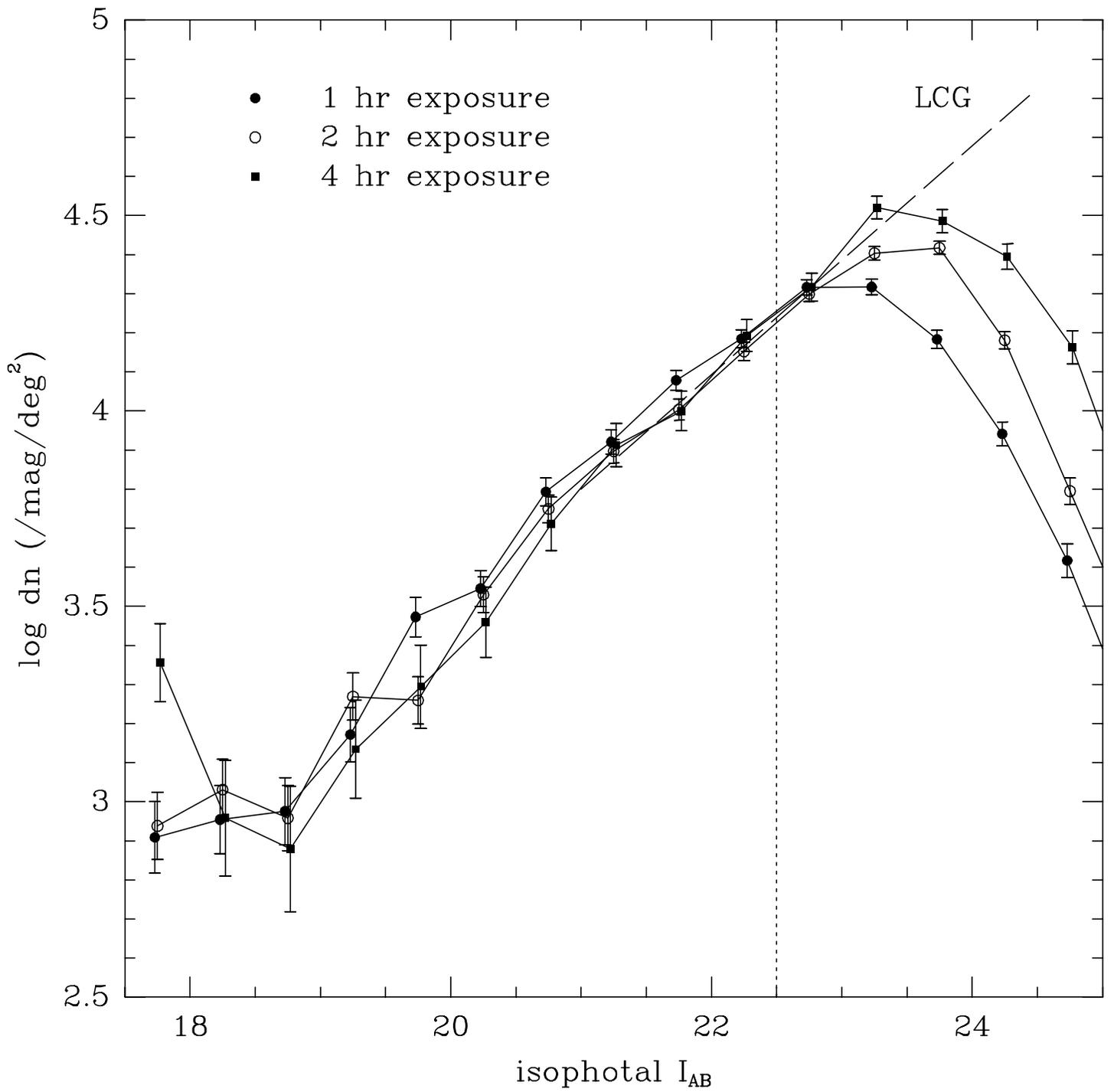

# The Canada-France Redshift Survey I: Introduction to the survey, photometric catalogs and surface brightness selection effects

S. J. Lilly[1]

Department of Astronomy, University of Toronto, Toronto, Canada

O. Le Fèvre[1]

DAEC, Observatoire de Paris-Meudon, 92195 Meudon, France

David Crampton[1]

Dominion Astrophysical Observatory, National Research Council of Canada, Victoria, Canada

F. Hammer[1] and L. Tresse[1]

DAEC, Observatoire de Paris-Meudon, 92195 Meudon, France

## ABSTRACT

The Canada-France Redshift Survey has been undertaken to provide a large well-defined sample of faint galaxies at high redshift in which the selection criteria match as closely as possible those of samples of nearby galaxies. The survey is designed to have a median redshift of z $\sim$ 0.6 corresponding to a look-back time of half the present age of the Universe for $\Omega \sim 1$. Such a survey can then be used for studying many different aspects of the evolution of galaxies over the interval $0 < z < 1$. In this paper we describe the selection of the fields, the multicolor imaging observations and the construction and validation of the photometric catalogs. Particular attention is paid to quantifying the unavoidable selection effects in surface brightness and their impact on the survey is assessed in the context of the properties of known populations of galaxies. The photometric catalogs contain several thousand objects brighter than $I_{AB} < 22.5$ and are essentially complete for central surface brightnesses as faint as $\mu_{AB}(I) \sim 24.5$ mag arcsec$^{-2}$. This should be sufficient to include both normal surface brightness galaxies and prototypes of extreme low surface brightness galaxies.

*Subject headings:* galaxies: evolution galaxies: photometry surveys

---

## 1. Introduction – scientific motivation

As is well known, telescopes have the remarkable ability, through light travel time effects, to generate images of the Universe at much earlier epochs, allowing us in principle to directly observe the evolution of populations of objects over cosmic time. For many years, we have had access to samples of quasars, radio galaxies and other more or less exotic objects at very high redshift, $z \gg 1$, for which the look-back time is a large fraction of the present-day age of the Universe. Although these objects may be intrinsically interesting, they are poorly understood and the unusual criteria used to find them makes their relationship to the general galaxy population highly uncertain. Unfortunately, the observational situation with more normal galaxies has been more fragmentary. Clusters of galaxies are known to $z \sim 1$ and cluster galaxies have been extensively studied out to $z < 0.5$, but since only a small fraction of galaxies are found in such extreme environments, these are unlikely to give information on truly typical galaxies, by which we simply mean those galaxies in which most of the stars in the Universe are located. Such galaxies are generally rather faint at high redshifts and, virtually by definition, have no distinguishing characteristics (such as strong radio emission, unusual colors, or location in extreme environments) by which they can be selected. Large samples of these galaxies can only be constructed through the systematic measurement of redshifts of large numbers of faint galaxies selected from deep images, and until recently, very little has been known about normal galaxies at cosmologically interesting redshifts.

Over the last few years, astronomers have started to explore the properties of normal galaxies in the distant Universe. A number of redshift surveys of field galaxies at successively greater depths have been carried out (see e.g. Broadhurst et al. 1988, Colless et al. 1990, Lilly et al. 1991, hereafter LCG, Cowie et al. 1991, Lilly 1993, Tresse et al. 1993, Colless et al. 1993, Songaila et al. 1994). The initial work was undertaken on samples selected in the B-band with the motivation of understanding the galaxy number counts which are steeper than most predictions for an unevolving galaxy population and which have therefore been interpreted as providing evidence for evolution of the galaxy population. A basic difficulty is that at significant redshifts this corresponds to selection in the ultraviolet. Unfortunately, there are still uncertainties regarding the ultraviolet properties of the local galaxy population and thus even the null- hypothesis "no-evolution" predictions can be disputed (see e.g. Koo et al. 1993). Samples selected at such short wavelengths are also likey to be dominated by the most active objects. Furthermore, the deeper samples were quite small (often just a few tens of objects) and comparisons of the data with evolutionary models was often made on the basis of crude statistics such as the medians of the redshift distributions.

The Canada-France Redshift Survey (CFRS) has been undertaken to improve this situation by generating a large sample of several hundred galaxies in which the selection criteria match, as far as possible, those criteria defining samples of local galaxies. The CFRS was formed by merging two independent $I$-band selected surveys that had been initiated on the 3.6-m Canada-France-Hawaii Telescope (CFHT), the initial results from which were reported by Lilly (1993) and Tresse et al. (1993). Upon merging these projects, we adopted the ambitious goal of acquiring a sample of 1000 objects selected to have $17.5 \leq I_{AB} \leq 22.5$ without regard to morphology, color or magnitude (within the above range). The faint limit of $I_{AB} \leq 22.5$, broadly equivalent to $B < 24$ or $K < 20$ in terms of projected number density on the sky, is the faintest practical limit for spectroscopy on 4-m class telescopes - redshifts can be secured with a high degree of completeness with integrations of order 8 hours. However, this depth is sufficient to yield a median redshift of $<z> \sim 0.6$ corresponding to look–back times of 50% of the Hubble time (for $\Omega \sim 1$) .

The specific scientific aims of this project were thus:

- To define the galaxy population in terms of the epoch- dependent luminosity function in the rest-frame B-band for different classes of galaxies over the interval $0 < z < 1$, with particular emphasis on $z \sim 0.6$ ($\tau/\tau_0 \sim 0.5$). At this redshift, the luminosity function was required to extend down to 0.3 $L^*$.

- To determine the small-scale clustering properties of the galaxy population over the same redshift interval, in order to track the growth of structure in the Universe and to understand the role of environment in driving the evolution of individual galaxies.

- To relate the faint field galaxy population to the population of faint radio sources revealed by deep VLA images and to serve as the basis for deep surveys in other wavebands.

- To allow the construction of well-defined sub-samples of galaxies at different epochs whose spectra and spectral energy distributions, morphologies, and kinematics could be studied so as to determine the physical processes occurring in individual galaxies that together produce the changes seen in the population as a whole.

These goals dictated several basic features of the survey. Firstly, the objects for spectroscopic study must be selected in the $I-$band which is equivalent to the rest-frame $B-$ and $V-$bands at the redshifts of most interest, i.e. the $I-$band is equivalent to the $V-$band at $z \sim 0.5$ and to the $B-$band at $z \sim 0.9$. Second, the spectroscopic observations must be carried out on all objects satisfying the basic magnitude selection criteria without regard to color or image morphology and compactness. While surface brightness selection effects are to a certain degree unavoidable, we required that these be quantifiable and minimal for known classes of galaxies. Selection in the K-band was considered (cf. Songaila et al. 1994) but rejected – although this allows selection at rest wavelengths longward of 1 $\mu$m, where the light is likely to be dominated by the oldest and most slowly evolving stellar populations, the small format of infrared array detectors and the high sky brightness would have precluded us from obtaining very deep images for a sufficiently wide area of sky, and it was felt that acquiring a large sample and minimizing surface brightness selection effects was of paramount importance. An additional practical difficulty is that spectroscopy of large samples of galaxies is at present possible only in the optical waveband ($\lambda < 1$ $\mu$m). Thus the objects which appear in a deep K-selected sample but not in the equivalent $I-$band sample are, by definition, extremely faint in the optical and it is extremely difficult to secure a spectroscopic identification for these objects.

This paper is the first in a series describing the CFRS. In this paper (CFRS I), we describe the selection of the survey fields, the initial imaging observations, and the construction of the photometric catalogs from which objects were selected for spectroscopic study. We also address the important question of surface brightness selection effects in the catalogs. The spectroscopic program is described in detail by Le Fèvre et al. (1995a, CFRS II) and additional aspects of the spectroscopic program are presented in Lilly et al. (1995a, CFRS III) and Hammer et al (1995a, CFRS IV). The spectroscopic and photometric data are integrated in Crampton et al. (1995, CFRS V) with the aim of verifying the spectroscopic identifications and constraining the natures and likely redshifts of the spectroscopically unidentifed objects. The global properties of the sample such as the color-redshift and magnitude-redshift diagrams as well as the overall N(z) histogram are also presented in CFRS V. The tri-variate luminosity function is constructed in Lilly et al. (1995b, CFRS VI), the nature of faint radio sources is discussed in Hammer et al (1995b, CFRS VII), and the computation of clustering parameters is carried out by Le Fèvre et al. (1995b, CFRS VIII). Further papers are planned to develop other aspects of the survey.

Unless stated to the contrary, values of $H_0 = 50$ km s$^{-1}$Mpc$^{-1}$ and $q_0 = 0.5$ are assumed. Magnitudes are quoted with the AB normalization scheme (Oke 1972). The offset from the $\alpha$-Lyrae-based normalization is as follows: $B_{AB} = B - 0.17$, $V_{AB} = V$, $I_{AB} = I + 0.48$, and $K_{AB} = K + 1.78$.

## 2. Selection of survey fields

The CFRS has been carried out in five survey fields of area $10' \times 10'$ chosen to match the field of view of the MOS spectrograph on CFHT. The five fields are listed in Table 1. The locations of the fields were chosen to be evenly spread in right ascension, for observing efficiency, and to be at moderate-to-high galactic latitude ($|b^{II}| > 45°$). Three of the fields are equatorial and two are in the north.

Most of the fields were located at major intersections of the [1950] coordinate grid and thus are random except that they were chosen to avoid bright stars and Zwicky (1966) clusters. One field was located in the area surveyed to very faint radio flux density limits by Fomalont et al. (1991) so that the relationship of faint radio sources to the field galaxy population could be studied (see CFRS VII) adn one of the fields contained the SSA-22 survey field initially set up by LCG and subsequently worked on by the Hawaii group (e.g. Cowie et al. 1991, Songaila et al. 1994), in order to allow for some external cross-checks.

## 3. Imaging data

The basic imaging data for the survey consists of high quality $V-$band and $I-$band CCD images obtained at prime focus on the 3.6m CFHT. In addition, $B-$band images were obtained for the 0000+00, 1415+52 and 2215+00 fields and infrared $K-$band images were obtained for most of the objects which were observed spectroscopically. A log of the imaging observing runs is given in Table 2.

### 3.1. Optical imaging

Optical imaging observations were carried out during a number of observing runs with the FOCAM camera on CFHT (see Table 2). For the V-band and $I-$band observations the "Lick-2" 2048x2048 CCD detector was used, while for the later B-band observations the "Loral-3" 2048x2048 chip, which had improved blue response, was used. In each case the plate scale was $0\overset{''}{.}207$ per pixel giving a field of view of about $7'$. In four of the five fields, a mosaic pattern of four separate images was obtained, offset in RA and declination by $3'$, in order to cover each $10' \times 10'$ field. Each of these four images was in turn made up of multiple exposures offset by a few arcsec to allow for the generation of sky-flats and the elimination of cosmic rays and chip defects. The final composite image of each survey field thus consists of a mosaic pattern in which the exposure time varies from the nominal amount in the four corners (36% of the image), to twice nominal in the cross-shaped overlap region (48% of the image) and to four times nominal in the central $4' \times 4'$ region (16% of the image). In the fifth field, 2215+00, the RA offset was about $6'$, leading to a larger survey area ($12' \times 10'$) but with reduced overlap. This "nominal" exposure time was generally 60 minutes for the primary V-band and $I-$band images, although in the 1415+52 field it was 45 minutes. The supplementary $B-$band images were shorter with a nominal 30 minute exposure time. The image quality was always sub-arcsecond and ranged between FWHM $0\overset{''}{.}6$ and $1\overset{''}{.}0$.

Reduction of the imaging data was straightforward and followed what are by now standard procedures. The cosmetic quality of the final images varied but was generally good to excellent. Photometric calibration was achieved through observations of equatorial standard stars from Landolt (1983) and the standard fields of Christian et al. (1985). These showed negligible color equations in $V$ and $I$ but a significant one in $B$, indicating that our effective wavelengths are 5500Å, 8320Å and 4570Å respectively. The photometric zero-point of all of the photometry was adjusted to be on the AB system (Oke 1972) using the offsets $B_{AB} = B - 0.17, V_{AB} = V$ and $I_{AB} = I + 0.48$

### 3.2. Infrared imaging

Infrared observations were made during two runs using the Redeye–Wide camera at cassegrain focus on CFHT (see Table 2). This instrument employs a 256x256 NICMOS-3 HgCdTe detector at $0\overset{''}{.}5$ per pixel giving a field of view of $2' \times 2'$. The smaller detector format of infrared arrays made it impractical to observe the full extent of each survey field in the infrared. Rather, since within each field the objects selected for spectroscopic study were located in three parallel strips separated by about 3.5 within each field (see CFRS II), we planned to obtain infrared images of only these three strips with the aim of obtaining infrared photometry on most of the objects for which we had spectroscopic information. Infrared observations were not obtained for the southern strip in the 0000+00 field, or for either the northern or southern strips in the 1415+52 field.

The observations consisted of a sequence of 30 second exposures between which the telescope was offset by approximately 12″ along the direction of the strip, giving about 10 exposures of each object. Usually three such sequences were generated for each of the strips of sky to be observed, with the offset used in each sequence being changed so as to minimize the effect of the well-known residual images that occur with these detectors. The filter used was a $K'$ filter (Wainscoat and Cowie 1991).

Reduction of these infrared data is more involved than that of the optical CCD data. Corrections were first applied for the non-linearity of the detectors. For each frame, a map of the varying background was generated from median averaging of the 8 adjacent images, and then subtracted. Residual patterns such as intermittent pickup noise were eliminated using a variety of median smoothing operations designed to preserve as much as possible the photometric integrity of the data. Nevertheless, systematic errors of up to 0.1 magnitude may be present in the infrared photometry.

Photometric calibration was achieved through observations of faint UKIRT standard stars from Casali and Hawarden (1992). The effective wavelength of the $K'$-band was taken to be 2.15 $\mu$m and, as with the optical data, the photometric zero-point was set on to the AB system by assuming $K'_{AB} = K' + 1.78$.

The $K'$-band strips were coregistered with the optical CCD fields using a 2nd-order transformation generated from approximately 20 objects visible on both sets of data.

## 4. Catalogue generation and photometry

The aim of the imaging observations was to produce a photometric catalog of all objects that approximates, as closely as possible, a simple magnitude-limited sample at $17.5 \leq I_{AB} \leq 22.5$ from which objects could be selected for spectroscopic study using the procedures described in CFRS II.

Object identification was carried out on the $I$–band images and the catalog was generated from isophotal magnitudes measured to a faint limiting isophote on these same $I$–band images. For each catalogued object, $(B)VI(K)$ colors were then generated using 3 arcsec aperture magnitudes measured from the images in the different wavebands.

In this section we describe the procedures used in constructing these catalogs and discuss the tests applied to validate them.

### 4.1. Image detection

Because of the origin of this project in two separate programs (Lilly et al 1993, Tresse et al 1993), two different image detection algorithms and isophotal magnitude programs were used to generate the basic catalogs of objects. These independent algorithms provide an important check on our procedures.

The catalogs for the 0000+00, 0300+00 and 1000+25 fields were generated by OLF using the FIND algorithm in DAOPHOT. The catalogs for the 1415+52 and 2215+00 fields were generated by SJL using the MULTIM algorithm in the GASP package. Both detection algorithms work on the usual principle of finding objects that consist of more than a minimum number of contiguous pixels lying above some threshold relative to a locally determined background, with criteria applied to separate close blends of objects. In both cases, the object detection is based on images that have been effectively smnoothed with a 3x3 boxcar. A preliminary catalog for the 1415+52 field was also generated by OLF from an independent set of images of this field with a shorter nominal exposure time of 15 minutes. This preliminary 1415+52 catalog, used by Tresse et al. (1993), was subsequently superseded by the one generated by SJL which was based on independent deeper images (45 minutes nominal exposure time). We compare these two catalogs below.

In order to be subsequently photometered, an object has first to be detected, and the completeness of this detection process as a function of total brightness and surface brightness is critical to the integrity of the catalog. Indeed, surface brightness selection effects in faint galaxy samples are potentially so important in faint galaxy research that we devote Section 5 of this paper to a discussion of this issue. The detection of objects is based primarily on their *central* surface brightnesses, since it is only in the inner regions of the images that the pixels contain enough signal to stand out above the sky background. In considering surface brightness selection effects, we therefore consider the central surface brightnesses of images (as modified by the effects of seeing), parameterizing this by the surface brightness (in mag arcsec$^{-2}$) of the central ($0\rlap.{''}207$ × $0\rlap.{''}207$) pixel of the image. For reference, a surface brightness of $\mu_{AB}(I) \sim 24.5$ mag arcsec$^{-2}$ produces a signal-to-noise ratio of unity *per pixel* on the shallower areas of the images (i.e. those with only the nominal exposure time).

We have investigated the completeness of our catalogs in three different ways. First, as discussed above, our mosaic images in 4 of the 5 fields consist of nine segments with exposure times varying by a factor of four (four at the nominal exposure time, four at twice nominal and the central one at four times nominal). We have examined the number counts of objects found in these four fields as a function of the exposure time, $t$. These are shown in Figure 1 along with a line representing the deeper counts of LCG. These latter are based on 6 hour exposures on CFHT with a more sensitive CCD and are consistent with the "corrected" $I$−band counts of Tyson (1988). As expected, the counts at faint magnitudes diverge for the different exposure levels and the depth of our data is clearly approximately proportional to $t^{1/2}$. However, at $I_{AB} < 23$, there is very close agreement between the counts at the different exposure levels. This suggests that even the shallowest parts of the images (with ~1 hour integration) are not missing a significant population of objects at $I_{AB} < 22.5$ that are detected on the deeper parts of the images. There is also good agreement bwteen our total counts and the LCG and Tyson (1988) data although we find variations of up to 20% in the counts at $I_{AB} = 22.5$ in the different fields. A variation of about 10projected correlation function w($\theta$) at this depth on 10 arcmin scales (see e.g. Hudon and Lilly 1995).

Second we have carried out tests in which the algorithm attempts to recover artificial objects added to the images and, again, have examined the results as a function of exposure time. The 1415+52 field was used for these experiments since it had the shortest nominal exposure time (45 minutes). The artificial objects all had a total magnitude of $I_{AB} = 22.5$, but had central surface brightnesses that ranged from $\mu_{AB}(I) = 22.5$ mag arcsec$^{-2}$ to $\mu_{AB}(I) = 25.5$ mag arcsec$^{-2}$ in 0.5 mag arcsec$^{-2}$ increments. We added 200 artificial objects at each level of central surface brightness. The results of this recovery exercise are shown in Figure 2. At all central surface brightnesses brighter than $\mu_{AB}(I) = 24.0$, more than 95% of the objects are recovered, independent of the exposure time. At fainter central surface brightnesses, an increasing fraction of objects are not recovered. The $t^{1/2}$ dependence on exposure time is also again evident. But, averaged over all regions of the image, over 80% of the objects with $\mu_{AB}(I) = 24.5$ are recovered, and given that the other survey fields have a 33% longer exposure time, we take this $\mu_{AB}(I) = 24.5$ central surface brightness level as a reasonable completeness limit for the catalogs. For the shallowest regions of the image, this corresponds to a $3\sigma$ detection in the 3x3 smoothing box. In fact, as is shown below (see Figure 7), rather few objects are either observed or expected to have such low central surface brightnesses at this magnitude level. This observed central surface brightness limit corresponds to rest-frame $\mu_{AB}(V) = 23.2$ at z = 0.5 and $\mu_{AB}(B) = 22.4$ at z = 0.9, although for compact objects the surface brightness will be dominated by seeing effects.

Finally we have compared the catalogs generated for the 1415+52 field by OLF and SJL from the independent algorithms applied to the two independent images of this field (see above). It should be remembered that the OLF catalog was based on images with only a 15 minute nominal exposure. The statistics of discrepancies (i.e. objects detected in one catalog but not in the other) are shown in Figure 3. At $I_{AB} < 22.5$ (i.e. the limit of the spectroscopic survey) the number of objects missing from one or other of the catalogs is about 5%. Examination of the images suggested that about 1/3 of these were spurious objects (especially objects near to bright stars) and 2/3 were real objects missed, for no obvious reason, by one or other of the programs. Losing small numbers of objects in this way appears to be a common feature of image detection algorithms. In the 1415+52 field, we inspected the discrepant objects and added those that had clearly been missed. In the other fields, we compared by eye the images and the lists of detected objects and added to the latter any obvious discrepancies before the photometry and final catalog construction was undertaken. In no case did the number of added objects exceed 3% of the total. At fainter magnitudes, the number of objects not found in the other catalog increases rapidly presumably both because of spurious objects and non- detections (it should be stressed in this context that the exposure time on the first 1415+52 image was quite short).

## 4.2. Isophotal photometry

Isophotal magnitudes were calculated to the $\mu_{AB}(I) = 28.0$ mag arcsec$^{-2}$ isophote. As noted above, this isophote has a very low S/N per pixel and can only be detected by azimuthal smoothing of the image, i.e., by constructing the radial profile of each image. In the rest-frame of distant galaxies, it corresponds to $\mu(V) = 26.7$ mag arcsec$^{-2}$ at z = 0.5 and $\mu_{AB}(B) = 26$ mag arcsec$^{-2}$ at z = 0.9 and thus would be expected to contain almost all the light.

We have calculated the effects of using different isophotal levels and different apertures for photometry on idealized seeing. The three curves on Figure 4 show the magnitude shortfall (relative to the total brightness of the galaxy integrated to infinity) as a function of the scale length ($h$ or $r_{eff}$) of the galaxy for three different photometric methods. The first, as in this work, is for an isophotal magnitude in which the isophotal limit (in mag arcsec$^{-2}$) is set 5.5 magnitudes below the total magnitude of the galaxy (in mag). The second is for an isophotal limit 3 magnitudes below the total magnitude (as used by Colless et al. (1993), albeit with a statistical correction of 0.15 mag evidently applied) and the third is for $3''$ aperture magnitudes (as used by for instance Cowie et al. 1991, 1994 and Steidel et al 1995). The scale lengths of typical faint galaxies are around $0''\!.4$ (Griffiths et al. 1994, Schade et al. 1995) so we would expect that (a) our isophotal magnitudes should closely approximate total magnitudes and (b) that other photometry schemes may underestimate the total light by around 0.2 of a magnitude. At a more empirical level, we show in Figure 5 the observed difference between our isophotal magnitudes and 3 arcsec aperture magnitudes as a function of magnitude for the non-stellar objects in the sample (the stars were excluded on the basis of their spectra - see CFRS V for a discussion). The mean offset at $I_{AB}\sim 22.5$ is about 0.27 magnitudes, with a large variation between objects at a given depth. This offset should be taken into account in comparing the results of different surveys.

The isophotal diameters of the objects at $I_{AB} \sim 22.5$ are typically about $5''$. A potential drawback of the faint isophotal levels and the correspondingly large isophotal diameters used in this work is that neighboring objects can distort the isophotes and be included in the isophotal aperture leading to substantial overestimates of the isophotal flux of the object. Both isophotal photometry schemes used in our photometry attempted to eliminate the effects of neighboring objects but were not always successful. This problem was dealt with after the spectroscopic observations had been made by examining the images of all the objects for which spectra had been taken. Any potential containments were then removed using the IMEDIT task in IRAF and a new isophotal magnitude computed. This procedure resulted in some objects moving below the $I_{AB} = 22.5$ limit of the statistically complete spectroscopic catalog. These objects were placed in the "supplemental" spectroscopic catalog (for which no claim of statistical completeness is made, see CFRS II).

As a final check on the isophotal photometry routines, we compared the photometry in the two catalogs for the 1415+52 field that were generated independently by our two packages from independent imaging data. This comparison is shown in Figure 6b. The bulk of the objects show a small scatter around the 45° line, and there is no significant systematic offset between the two measurements on each object down to the faintest levels (Figure 6a). A few outliers were found to be substantially brighter in one or other catalog. Examination of these outliers on the images showed that these were always due to companions entering the isophotal aperture and, as noted above, this problem was subsequently corrected for the objects observed in the spectroscopic program. The scatter in the photometry as a function of magnitude is shown in Figure 6a. The r.m.s. magnitude difference is roughly constant at bright magnitudes and increases at the fainter magnitudes to 0.3 mag at $I_{AB} \sim 22.5$. This was modelled as the sum of a constant and a magnitude dependent term (see Figure 6a). Given that the exposure times for these data were only 75% and 25% of the exposure times used in the remainder of the survey, we estimate that the photometric uncertainty in the isophotal magnitudes in the bulk of the survey is typically 0.17 mag at the $I_{AB} \sim 22.5$ limit of the objects selected for spectroscopy.

### 4.3. Aperture Photometry

Aperture magnitudes were generated from $3''$ diameter apertures for each available $B, V, I$ and $K'$ image. Uncertainties for the different exposure levels around the mosaic pattern were calculated by randomly placing these apertures in neighboring areas on the images and calculating the standard deviation of the sky values.

### 4.4. Compactness parameter

The compactness of all detected images was parameterized using the Q parameter of Le Fèvre et al. (1986). This is compared to other compactness parameters by Slezak (1988). In essence, this parameter represents the inverse of the ratio of central surface brightnesses of the object in question and a star of the same isophotal magnitude. The Q parameter could in principle be used as a star-galaxy discrimnator, although in our spectroscopic program all objects were observed regardless of their compactness. We discuss in CFRS V the degree to which the spectroscopic identification correlates with the Q parameter.

### 4.5. Astrometry

Astrometric positions (equinox 2000.0) of all objects were determined relative to stars in the Hubble Space Telescope Guide Star Catalogue. In the 1415+52 and 2215+00 fields this was done via measurements of non-saturated stars visible on the Palomar Observatory Sky Survey prints. In the remaining fields, the positions of the stars in the Guide Star Catalogue were used directly. An external check is provided by the radio sources in the 1415+52 field (Fomalont et al. 1991) which show a systematic offset of less than $0''\!.2$.

## 5. Surface brightness selection effects

There has been considerable debate recently concerning the nature of surface brightness selection effects in deep redshift surveys such as the CFRS. Surface brightness effects can both reduce the fraction of the light of a galaxy contained within an isophote (see section 4.2 above) and can lead to galaxies being missed altogether in the survey (see section 4.1). As discussed above, we believe that our sample is essentially complete for all galaxies which have (after the effects of $0''\!.9$ seeing) central $\mu_{AB}(I) = 24.5$ and that our isophotal photometry to $\mu_{AB}(I) = 28$ includes more than 90% of the light of faint objects. In this section we seek to place this surface brightness limit in context.

Two quite different arguments have been made in the literature regarding surface brightness selection effects in deep redshift surveys such as these. On the one hand, it has been argued (Yoshii 1993) that many deep samples are *biased* against high redshift objects because of the well known $(1+z)^4$ surface brightness dimming with redshift. This has been invoked to explain the absence of high redshift galaxies (z > 0.7) in the published B-selected samples of Colless et al. (1990) and Cowie et al (1991). On the other hand, McGaugh (1994) has argued that in fact the faint surveys, generated from deep CCD images, may *include* many low surface brightness galaxies that would have been missed in brighter surveys that used photographic plate material for the initial object detection. This idea has been developed by Ferguson and McGaugh (1994) who have constructed heuristic models illustrating how the local luminosity function of Loveday et al. (1992) may substantially underestimate the slope of the faint end of the local luminosity function.

Unfortunately, observers have often inadequately described the surface brightness selection criteria that are inevitably present in all galaxy samples. In particular, distinction has not always been made between the isophotal level for photometry, and the isophotal level for initial detection. As illustrated above, the former is usually much fainter than the latter since detection must be based on the high signal to noise peaks whereas the photometric signal must be integrated out to large radii, where the signal to noise per pixel is very small, if the flux from each object is not to be systematically underestimated.

Figure 7 shows the central surface brightness as a function of isophotal magnitude for the present sample. We show only those objects observed spectroscopically, since it is only for these that the effects of any companions on the isophotal magnitude have been reliably dealt with. A small adjustment has been made to bring the stellar loci in each field into alignment, and the diagram effectively represents a uniform seeing of about $0''\!.9$. This was chosen partly because it is the seeing on our worst images and also because it has the convenient property that the central surface brightness (in mag arcsec$^{-2}$) for stars is approximately equal to the integrated magnitude (in mag). The vertical lines mark the flux density limits for the sample, $17.5 \leq I_{AB} \leq 22.5$, and the horizontal line marks the $\mu_{AB}(I) \sim 24.5$ central surface brightness limit discussed above. It can be seen that the majority of objects, even at the sample limit of $I_{AB} \sim 22.5$ have central surface brightnesses that are well above the nominal limit of $\mu_{AB}(I) \sim 24.5$ and, in a purely empirical way, the sample on Figure 7 gives a strong impression of being defined by the vertical lines, as desired if we are to have a simple flux density limited sample.

Figure 7 also shows the expected tracks, as a function of redshift, of a number of "normal" galaxies, calculated using the spectral energy distributions of Coleman, Wu and Weedman (1980). The tracks shown are for (a) an L* face-on exponential disk with the "Freeman" central surface brightness $\mu_{0,AB}(B) = 21.5$ (see Phillips et al. 1987 and van der Kruit 1987), $h = 6.6$ kpc and $M_{AB}(B) = -21.0$, (a') the same disk seen inclined with an axial ratio of 3:1, (b) and (b') face-on and inclined 0.1L* disks with the same $\mu_{0,AB}(B) = 21.5$ but with $h = 2.0$ kpc and $M_{AB}(B) = -17.5$, (c) an L* spheroid with $r_{eff} = 6.0$ kpc and $M_{AB}(B) = -21.0$ and (d) a 0.1L* spheroid with $r_{eff} = 1.4$ kpc. The points along each track represent (left to right) redshifts of 0.1, 0.25, 0.5, 0.75 and 1.0. Note that at the faint end of the sample the tracks are essentially parallel to the stellar locus, with $\mu_{AB}(I) \sim I_{AB}$ + constant, implying a constant effective image size. These parallel tracks arise because of two effects: the galaxies are either sufficiently small that seeing dominates or, if larger, the galaxies are more luminous and thus seen at sufficiently high redshifts so that the angular size is no longer decreasing as the redshift, and hence integrated magnitude, increases.

The distribution of points relative to these parallel tracks in the last two magnitudes of the survey are reassuringly constant, i.e., at $I_{AB} \sim 20.5$, the proportion of objects with $\mu_{AB}(I) > 22$ seems to be similar to that two magnitudes fainter at $I_{AB} \sim 22.5$ with $\mu_{AB}(I) > 24$. This is shown in Figure 8a which shows the histograms of $\mu_{AB}(I)$-$I_{AB}$, (i.e. of the vertical displacement from the diagonal line in Figure 7) for objects in different magnitude intervals. The width of these histograms gets gradually smaller to fainter magnitudes. Figure 8b shows the mean displacement of the galaxies in Figure 7. The gradual reduction in this quantity is expected on the basis of the theoretical tracks on Figure 7 and the absence of any abrupt changes to the distribution of $\mu_{AB}(I)$-$I_{AB}$ at the $I_{AB}$ limit of the survey, further support the idea that most galaxies exit the sample by passing through the magnitude selection rather than by surface brightness selection effects.

We have shown on Figure 9 the tracks of two quite different galaxies with low surface brightness characteristics. First we show a galaxy with a disk component with $h = 32$ kpc and $M_{AB}(B) = -22.6$ and a spheroidal component with $r_{eff} = 5.0$ kpc and $M_{AB}(B) = -21.5$, these parameters being chosen to be representative of F568- 6 and 1226+0105, two of the prototypical "Malin" giant low surface brightness galaxies discussed by Sprayberry et al. (1993). We have plotted the disk component alone, the bulge component alone and the two together. Note how, even though the *disk* of this galaxy would not appear in the sample (at least at most redshifts) because of its low central surface brightness, the *bulge* is readily detectable to z ~ 1. Furthermore, once the bulge is detected, the isophotal photometry should include most of the disk light (as indicated by the horizontal offset between the bulge and total tracks in the figure) and even this extreme object, viewed as a whole, leaves the sample at high redshift through the vertical flux density criterion, as desired.

Second, we show a 0.1 $L^*$ disk galaxy with central surface brightness of $\mu_{0,AB}(B) = 24.0$, i.e. 2.5 magnitudes below the Freeman value. This galaxy is representative of a member of the extreme "constant-size" Model B galaxy population of Ferguson and McGaugh (1995). This galaxy would probably be lost through surface brightness criteria at z ~ 0.25 whereas the more compact 0.1 $L^*$ Freeman disk (the track (b) in Figure 7) is present in the magnitude limited sample to z ~ 0.35. However, the density of objects on Figure 8 in the neighborhood of this track is very much lower than near the tracks of higher surface brightness galaxies, and the comoving density of any such objects must be low.

Thus, although there must inevitably be surface brightness selection criteria operating at some level in our catalogs, we believe that our imaging data are sufficiently deep that our galaxy catalog between $17.5 \leq I_{AB} \leq 22.5$ is to all intents and purposes limited by isophotal flux density alone and, furthermore, that the isophotal levels used in the photometry are sufficiently faint that they approximate "total" measurements. We believe that this is true both for "normal" galaxies and for the prototypes of many of the most extreme low surface brightness galaxies that have been found to date.

## 6. Color distributions

The distributions of $(V - I)_{AB}$ and $(I - K)_{AB}$ colors as a function of isophotal $I_{AB}$ magnitude for all objects in the photometric catalogs are shown for reference in Figure 10 together with median values which are insensitive to the presence of upper or lower limits. The median colors change slowly with magnitude in this range, which, it should be noted, is somewhat brighter than the onset of the rapid bluening trend seen in very faint galaxy samples (see e.g., Lilly et al. 1991 and Cowie et al. 1995).

$(B - I)_{AB}/(I - K)_{AB}$ and $(V - I)_{AB}/(I - K)_{AB}$ color-color diagrams are presented in CFRS V for the objects studied spectroscopically, where they are used to both check the consistency of the spectroscopic identifications with the photometric colors and to constrain the natures of objects for which no spectroscopic identification was secured.

## 7. Conclusions

The Canada-France Redshift Survey has been undertaken to study a large number of faint galaxies in the 0.1 < z < 1 range, with particular emphasis on the z ~ 0.6 region which corresponds

– 14 –

to look back times of $0.5\tau_0$ for $\Omega \sim 1$. In this paper, we have described the imaging observations in $B, V, I$ and $K$ and the subsequent construction of the photometric catalogs from which objects were selected, on the basis of their isophotal $I_{AB}$ magnitudes, for spectroscopic study. The use of isophotal magnitudes computed to a faint isophotal level ($\mu_{AB}(I) \sim 28$) means that our magnitudes closely approximate total magnitudes. The depth of our original images means that the sample is still 80% complete for objects with a central surface brightness as low as $\mu_{AB}(I) \sim 24.5$. This is sufficient to ensure that, for objects $I_{AB} \leq 22.5$, our catalogs are limited to all intents and purposes only by isophotal magnitude, both for normal high surface brightness galaxies and also for many prototypes of low surface brightness galaxies.

We particularly wish to thank the Directors of CFHT during the course of this project, Guy Monnet and Pierre Couturier, for their enthusiastic support of the CFRS project. We are also grateful to our respective national TACs for their allocations of telescope time to this collaborative project. We appreciate the referee's patient and careful reading of the first six manuscripts in this series. SJL's research is supported by NSERC of Canada, and we acknowledge additional travel support from NATO.

Fig. 1.— Number counts of objects (from the 0000+00, 0300+00, 1000+25 and 1415+52 fields) constructed from regions with exposure times of order $1^h$, $2^h$ and $4^h$. The dashed line represents the deeper counts of Lilly et al. (1993). In the present work, the extra exposure time results, as expected, in a deeper sample, but the effects of this are only apparent below the $I_{AB} = 22.5$ limit of the sample observed spectroscopically and the counts at $I_{AB} = 22.5$ are independent of the exposure time, suggesting that the shallower data is not missing a signficant population of objects detected in the deeper data.

Fig. 2.— The percentage of artificial objects (all with $I_{AB} = 22.5$) that were recovered by the image detection algorithm from regions with different exposure times in the 1415+52 field, as a function of the central surface brightness. At each level and in each region, 200 artificial objects were added. The small crosses show the weighted average of all regions. Over 80% of objects were recovered for central $\mu_{AB}(I) = 24.5$.

Fig. 3.— Histograms showing the number of objects in common (open area) between the two independent catalogs generated from two independent images of the 1415+52 field and the numbers detected in only one of the catalogs as a function of isophotal $I_{AB}$. About 1/3 of the missed discrepant objects are spurious, the remainder were missed by one or other deterction algorithm. It should be noted that these images had shorter exposure times than the remainder of the survey (45 minutes and 15 minutes as against 60 minutes). The number of objects in common decreases significantly below the $I_{AB} = 22.5$ limit.

Fig. 4.— The loss of flux (relative to the "total" light) of various photometry schemes as a function of the galaxy scale length for idealized exponential disks and de Vaucouleurs spheroids observed in $0\rlap{.}''9$ seeing. The three photometry schemes shown are (a) isophotal with $\mu_{lim} = m_{tot} + 5.5$ (as used in the present work), (b) isophotal with $\mu_{lim} = m_{tot} + 3$ and (c) aperture with $3''$ diameter. See text for discussion and details.

Fig. 5.— Variation of the average and r.m.s. magnitude difference between the isophotal magnitudes and the $3''$ aperture magnitudes in the sample as a function of magnitude.

Fig. 6.— (a) Lower panel. Comparison of isophotal photometry in the two independent catalogs generated from independent images of the 1415+52 field. As in Figure 3, it should be noted that these images were shallower than in the remainder of the survey. The outliers are due to contamination by neighboring objects. All objects observed spectroscopically were examined and if required re- photometered to eliminate this problem. (b) Upper panel. Systematic and random differences in magnitudes from (a). Points show medians of sets of 51 objects, vertical bars show the r.m.s. with outliers removed. The light dashed lines shows a model consisting of magnitude dependent and magnitude independent terms and the heavy dashed line shows the implied uncertainty *per observation* in the sample.

Fig. 7.— Central surface brightness as a function of isophotal magnitude for the objects observed spectroscopically. The vertical lines represent the sample limits of $17.5 \leq I_{AB} \leq 22.5$ and the horizontal dashed line indicates the limiting central surface brightness of $\mu_{AB}(I) \sim 24.5$ (cf. Figure 2). An ideal magnitude-limited sample would be limited only by the vertical lines. The tracks indicate the expected behaviour of "normal" galaxies with (a) an L* face on exponential disk with $\mu_{0,AB}(B) = 21.5$, $h = 6.6$ kpc and $M_{AB}(B) = -21.0$, (a') the same disk seen inclined with an axial ratio of 3:1, (b) face-on 0.1L* disk with $\mu_{0,AB}(B) = 21.5$, $h = 2.0$ kpc and $M_{AB}(B) = -17.5$, (b') this disk inclined at 3:1 axial ratio, (c) an L* spheroid with $r_{eff} = 6.0$ kpc and $M_{AB}(B) = -21.0$ and (d) a 0.1L* spheroid with $r_{eff} = 1.4$ kpc. The points along each track represent (left to right) redshifts of 0.1, 0.25, 0.5, 0.75 and 1.0. The distribution of the data points and the expectations based on these tracks suggest that indeed the sample is to all intents and purposes free of significant surface brightness selection effects.

Fig. 8.— (a) The trend of the mean (and error of the mean) displacement in the distribution of central surface brightnesses of objects relative to that of stars of the same isophotal magnitude (which have $\mu_{cen} \sim I_{iso}$ in $0''\!.9$ seeing - see Figure 7) for objects observed spectroscopically in three different magnitude ranges. There is no sudden decrease as the sample limit is approached. (b) Histogram representation of (a). At bright magnitudes the stars form a peak at zero which diminishes in size to fainter magnitudes. These histograms show a gradual increase in the compactness of the galaxies, as expected from the tracks in Figure 7 and no significant loss of low central surface brightness galaxies in the last 0.5 magnitude of the $I_{AB} \leq 22.5$ sample studied spectroscopically.

Fig. 9.— As for Figure 7, but showing a disk of $h = 32$ kpc and $M_{AB}(B) = -22.6$, a spheroid with $r_{eff} = 5.0$ kpc and $M_{AB}(B) = -21.5$, and a composite galaxy consisting of the sum of these two components. These parameters were chosen to be representative of F568-6 and 1226+0105, two of the prototypical "Malin" giant low surface brightness galaxies. Although the disk would be unlikely to be detected on its own, the bulge is readily detectable to z ~ 1 and the isophotal magnitude will include most of the light of the extended low surface brightness disk. The case of a 0.1 L* disk with scale length 6.6 kpc, representative of the extreme constant size Model B galaxy population of Ferguson and McGaugh (1995) is also shown as solid squares labelled FM95.

Fig. 10.— (a) 3 arcsec aperture $(V - I)_{AB}$ colors as a function of isophotal $I_{AB}$ magnitude for the entire photometric sample. Upper limits are plotted as for detections and this causes the truncation of the distribution at red coliurs. The line and large symbols trace the median colors. (b) as for (a) except for the $(I - K)_{AB}$ color.

*Table 1  CFRS survey fields*

| Field | RA [2000] | dec[2000] | $b_{II}$ | $l_{II}$ |
|---|---|---|---|---|
| 0000+00 | 00 02 40 | -00 41 45 | -60 | 95 |
| 0300+00 | 03 02 40 | +00 10 21 | -48 | 179 |
| 1000+25 | 10 00 44 | +25 14 05 | +53 | 208 |
| 1415+52 | 14 17 54 | +52 30 31 | +60 | 97 |
| 2215+00 | 22 17 48 | +00 17 13 | +46 | 63 |

*Table 2  Imaging observations*

| Dates | Wavebands (Instrument) | Pixel scale | Fields observed |
|---|---|---|---|
| Jul 7-8 1991 | I (FOCAM) | 0.207 arcsec | 2215+00 |
| Nov 6-8 1991 | V,I (FOCAM) | 0.207 arcsec | 0000+00, 0300+00, 1000+15 |
| Jun 2-3 1992 | V, I (FOCAM) | 0.207 arcsec | 1415+52, 2215+00 |
| Feb 3-6 1993 | K' (Redeye) | 0.50 arcsec | 0300+00, 1000+15, 1415+52 |
| Jul 13-14 1993 | B (FOCAM) | 0.207 arcsec | 0000+00, 1415+52, 2215+00 |
| Oct 20-21 1993 | K' (Redeye) | 0.50 arcsec | 2215+00, 0000+00, 0300+00 |

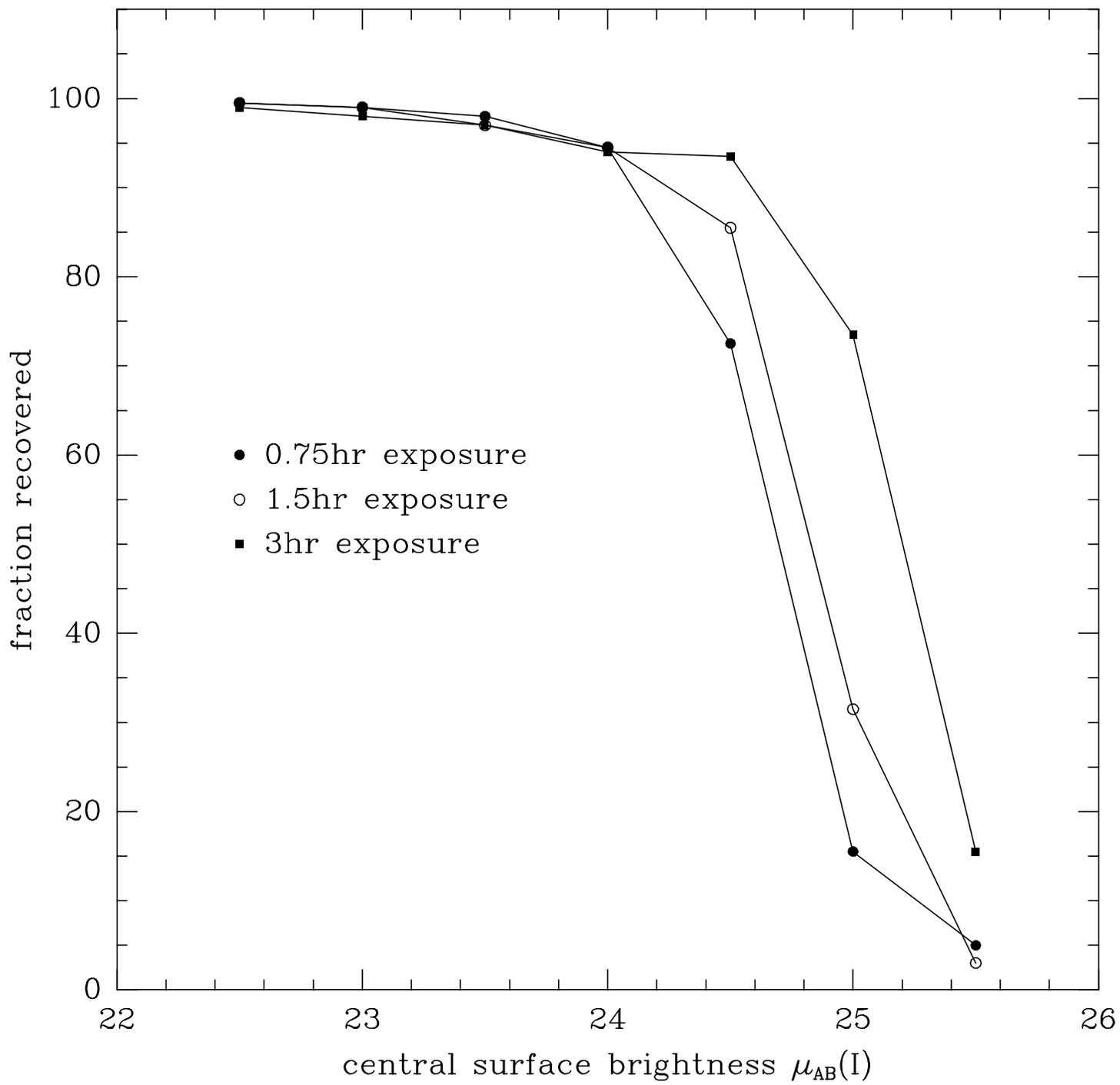

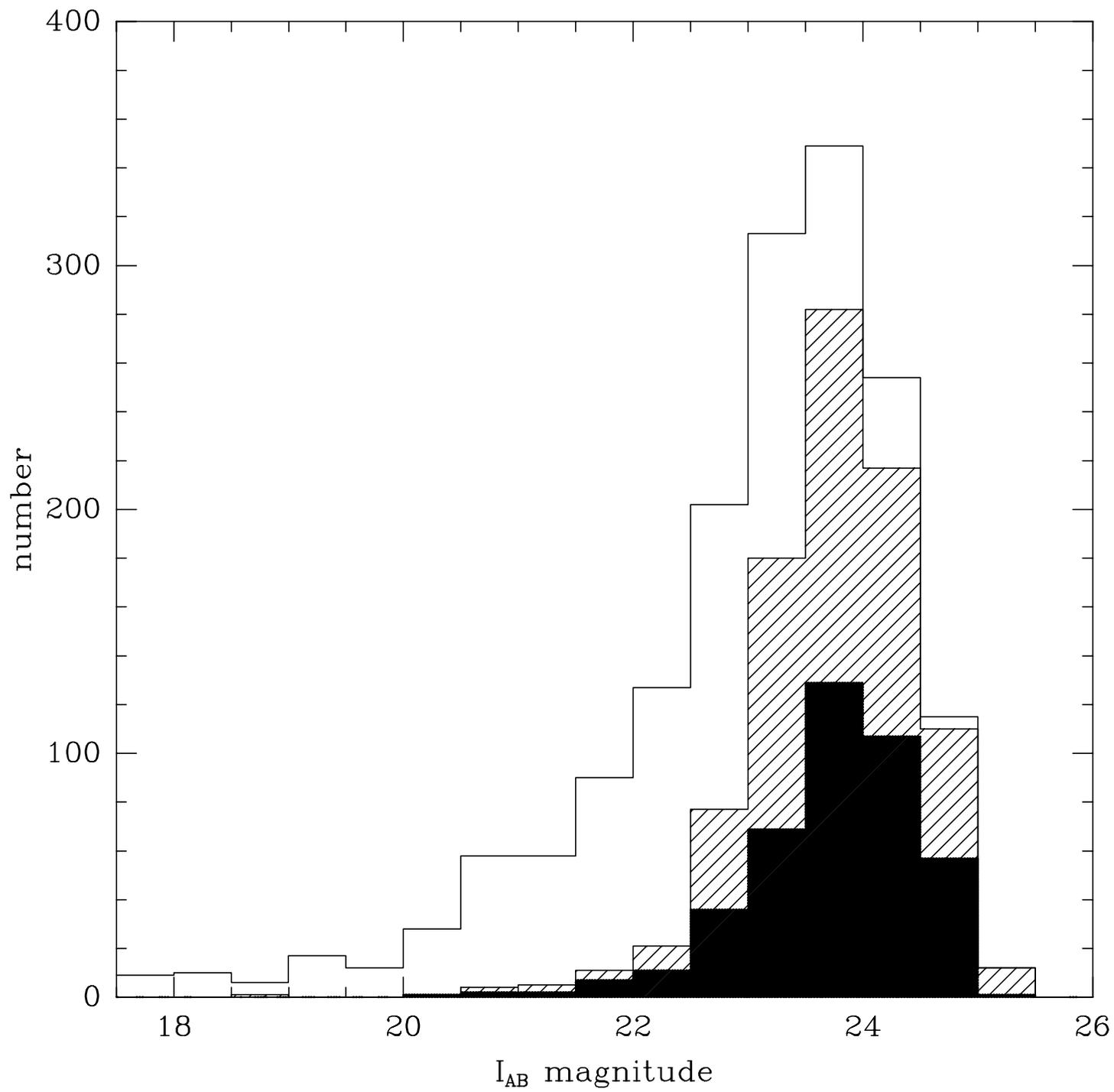

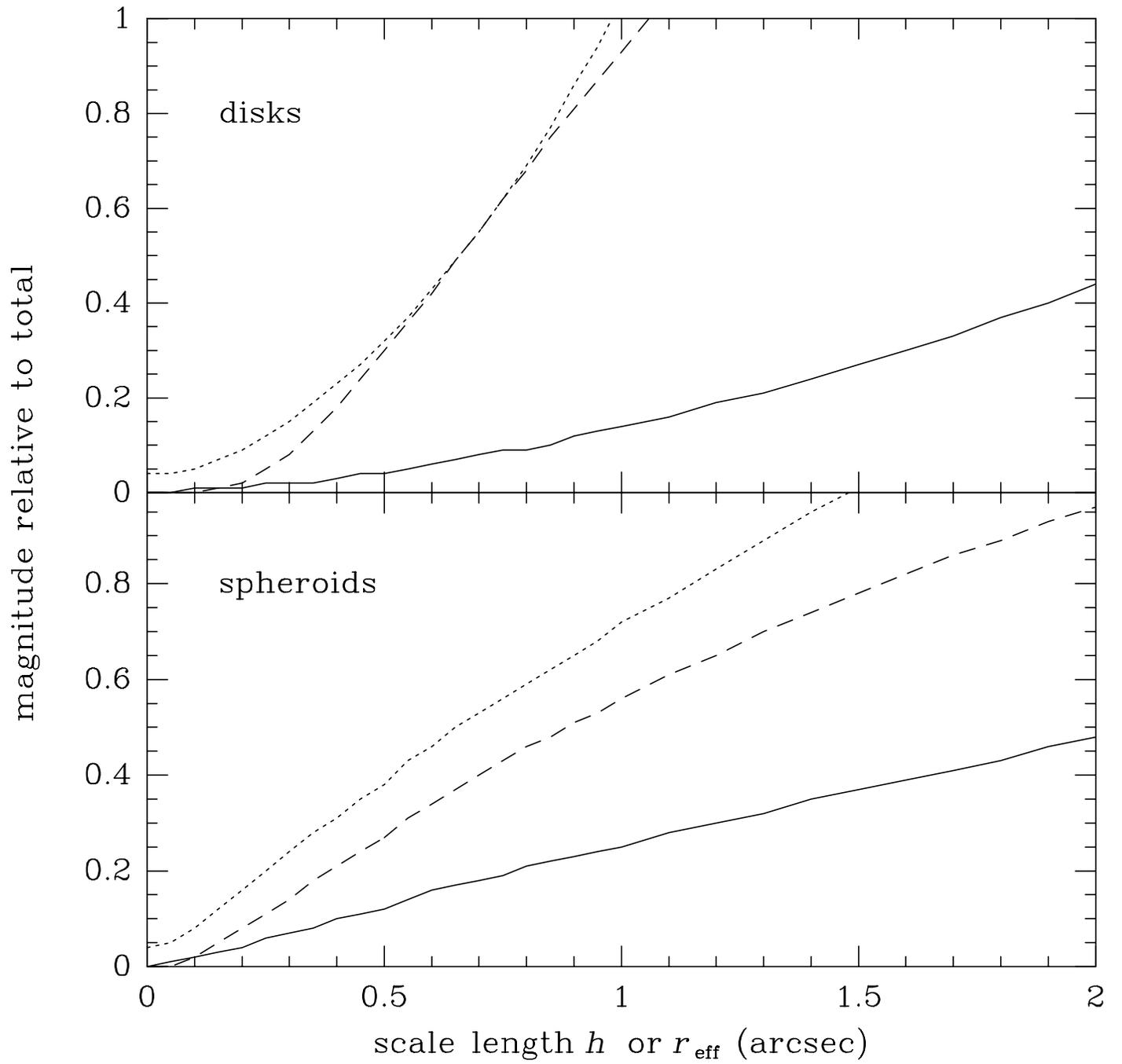

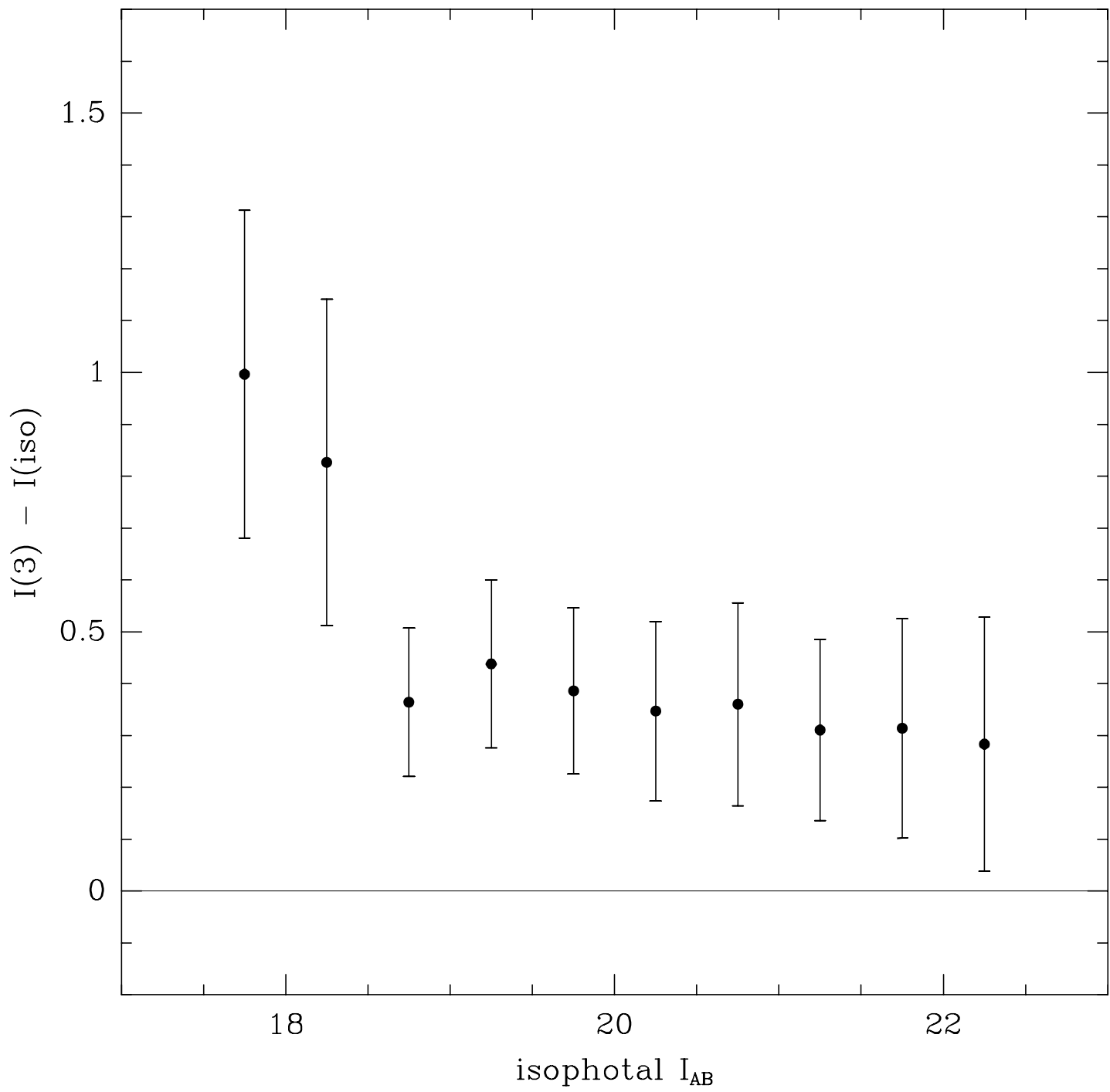

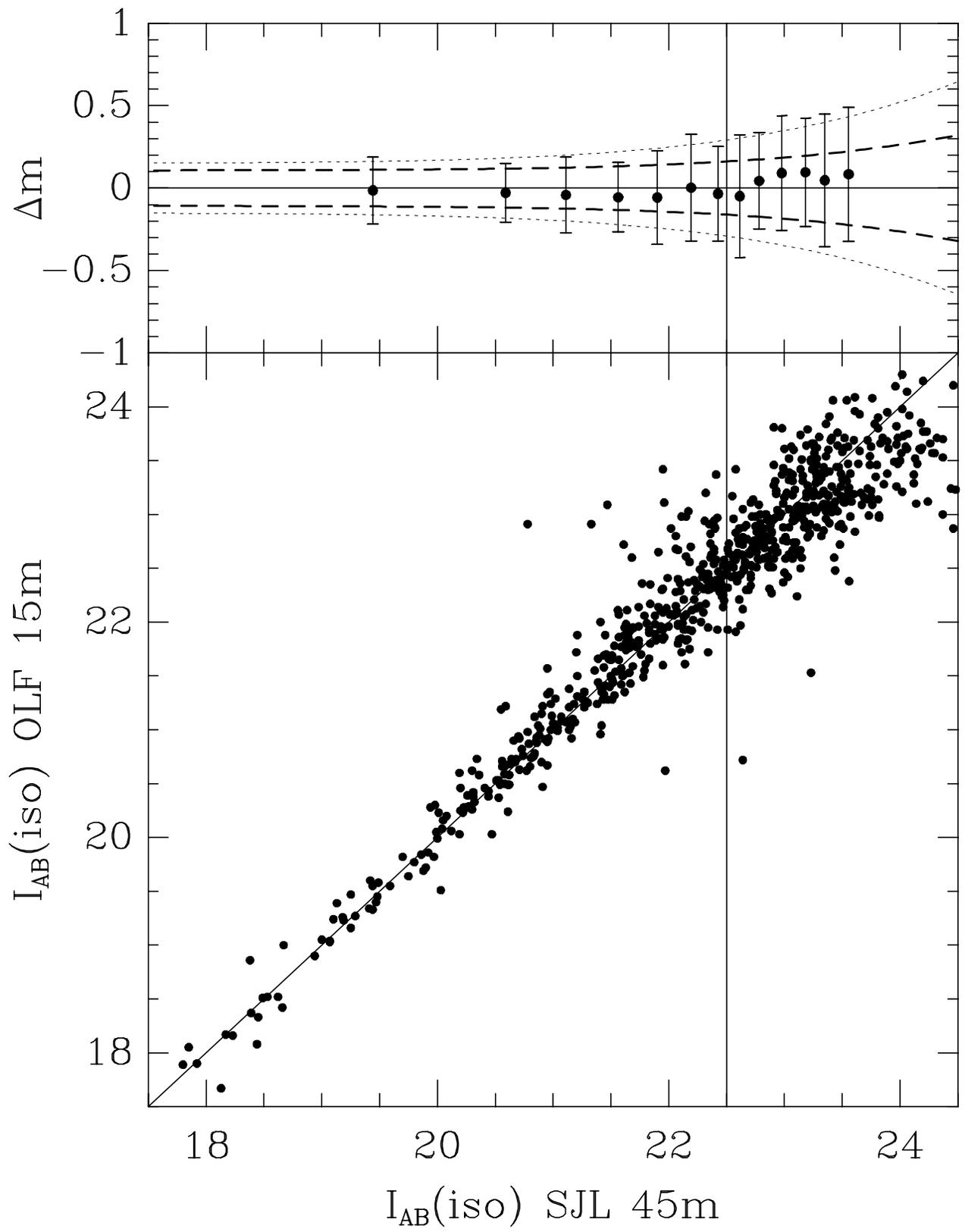

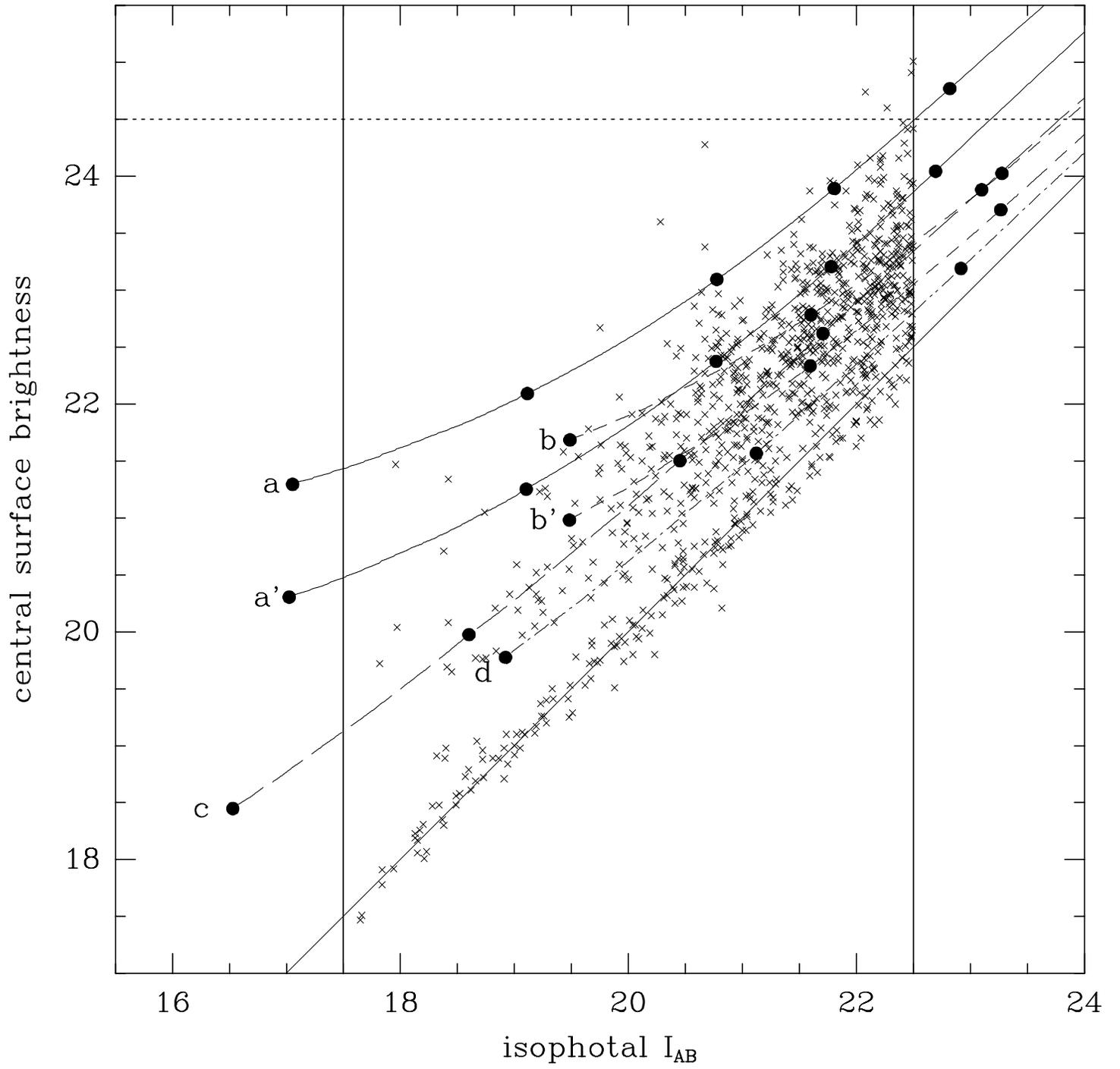

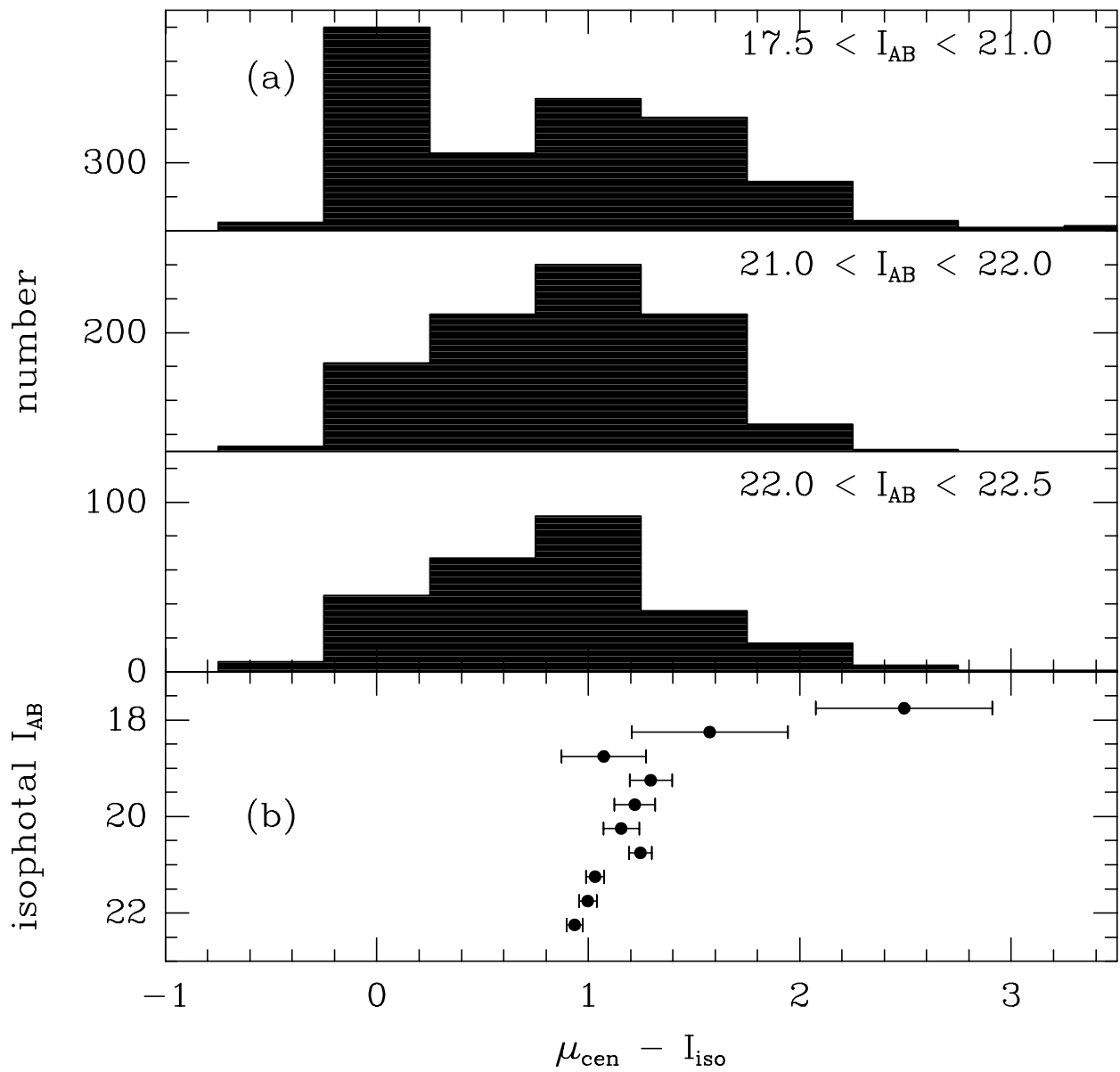

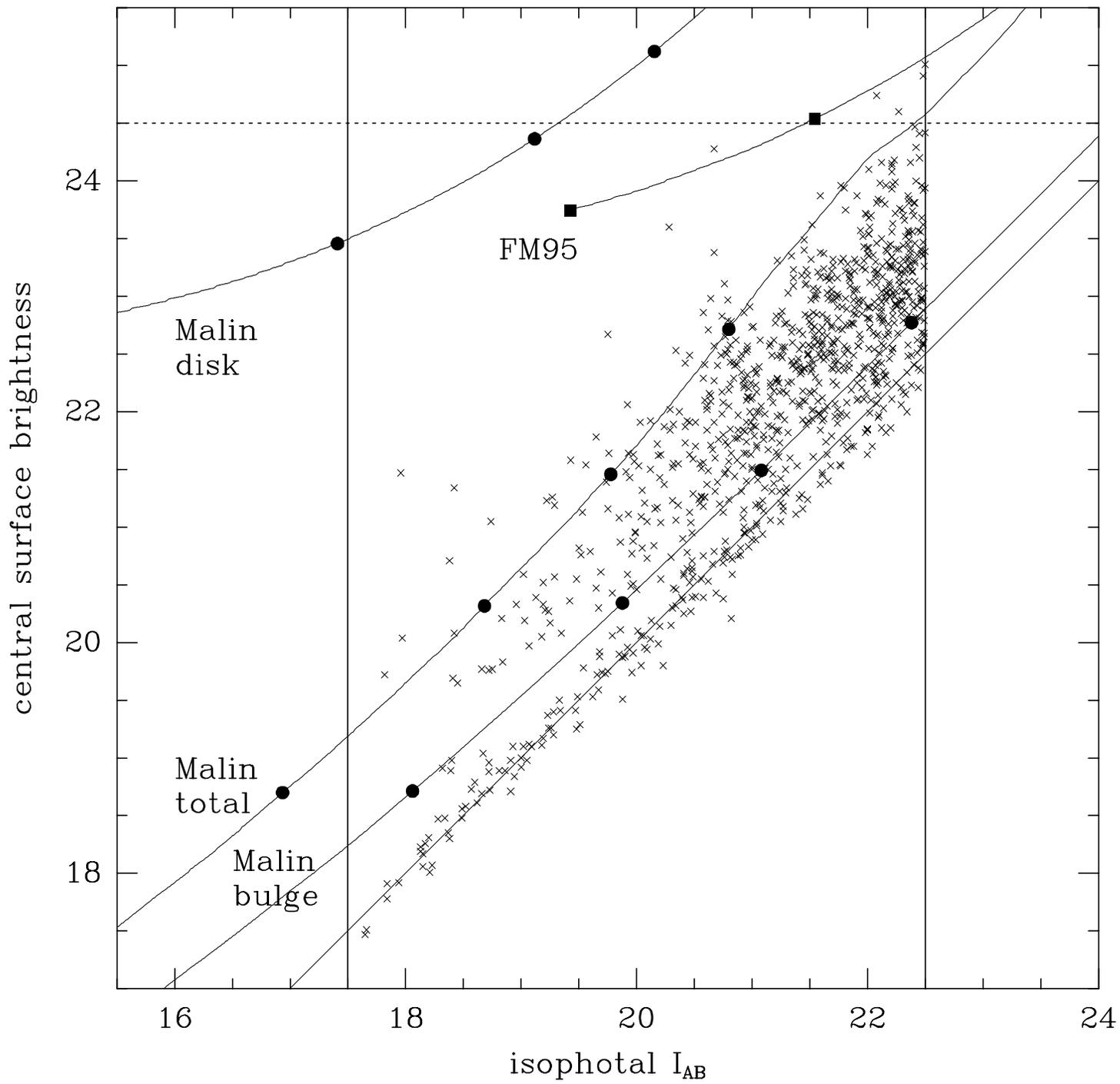

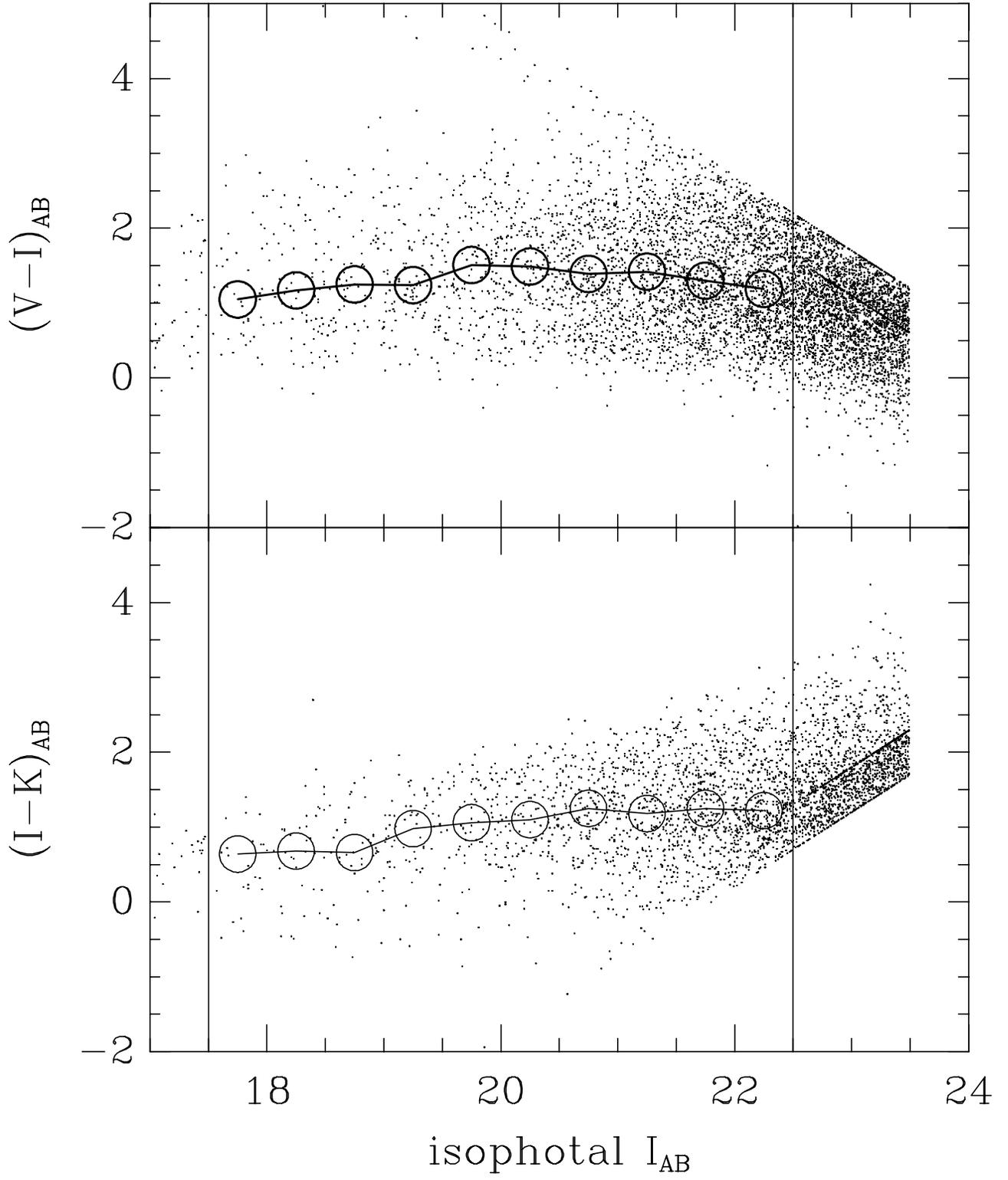